\documentclass{sig-alternate-05-2015}

\usepackage{caption}
\usepackage{subcaption}
\usepackage{graphicx}

\usepackage{listings}
\usepackage{balance}  
\usepackage{txfonts}
\usepackage{times}    

\usepackage{color}
\usepackage{textcomp}
\usepackage{booktabs}
\usepackage{ccicons}
\usepackage{todonotes}
\usepackage{nameref}

\usepackage{url}      

\makeatletter
\def\@copyrightspace{\relax}
\makeatother

\begin{document}






%

\title{A Warm Welcome Matters! The Link Between Social Feedback and Weight Loss in /r/loseit\titlenote{This is a preprint of an article appearing at WWW 2017}}
%
%
%
%
%


\author{Tiago O. Cunha,$^{1,2}$ Ingmar Weber,$^1$ Gisele L. Pappa$^2$ \\
\\
\affaddr $^1$Qatar Computing Research Institute, HBKU, Qatar\\
\affaddr $^2$Federal University of Minas Gerais, Brazil\\
\affaddr tocunha@dcc.ufmg.br, iweber@qf.org.qa, glpappa@dcc.ufmg.br 
}



\maketitle
\begin{abstract}
Social feedback has long been recognized as an important element of successful health-related behavior change. However, most of the existing studies look at the effect that offline social feedback has. This paper fills gaps in the literature by proposing a framework to study the causal effect that receiving social support in the form of comments in an online weight loss community has on (i) the probability of the user to return to the forum, and, more importantly, on (ii) the weight loss reported by the user. Using a matching approach for causal inference we observe a difference of 9 lbs lost between users who do or do not receive comments. Surprisingly, this effect is mediated by neither an increase in lifetime in the community nor by an increased activity level of the user. Our results show the importance that a ``warm welcome'' has when using online support forums to achieve health outcomes.
\end{abstract}

%
%
\printccsdesc



\section{Introduction}\label{sec:introduction}

Over the last decades, obesity rates have increased in many countries around the world, making the condition a major public health problem. Obesity is associated with significantly increased risk of more than 20 chronic diseases and health conditions \cite{thiese:2015}, and directly affects quality of life. In the U.S.\ 68.8\% of adults are overweight or obese\footnote{\url{http://www.niddk.nih.gov/health-information/health-statistics/Pages/overweight-obesity-statistics.aspx}}. Financially, health issues related to obesity and life style diseases impose an ever-increasing burden with medical costs linked to obesity estimated at USD 147 billion in 2008\footnote{\url{http://www.cdc.gov/chronicdisease/overview/\#sec3}}.



Though the percentage of U.S.\ adults self-reporting to be on a diet in any given week has fallen from 31\% in 1991, 20\% are still trying to lose weight through dieting at any point in time\footnote{\url{http://goo.gl/0naf5U}}. Previous studies have tied successful weight loss as well as other positive health outcomes to the presences of social support. In particular for long-term behavioral changes required for achieving and maintaining weight loss, user engagement is essential \cite{teixeira:2015}. Previous studies showed that users who stay longer in these programs have greater success in achieving their goals \cite{patrick:2011}.

%

Through the internet and social media it has become easier for users to find virtual support groups for anything from weight loss to drug addiction or depression. Though there are a number of studies looking at the effect of receiving social support on sustained engagement with an online community \cite{Burke:2009,cunha:2016, backstrom:2008}, the effect of such support on health outcomes has not been thoroughly studied. In particular, it is not clear how large one would expect the effect of receiving online comments on observed health outcomes to be as, arguably, many other effects from one's social environment should dominate the benefits of receiving encouragement from potential strangers online.

To study the importance of online support, one would, ideally, set up a proper experiment with a randomized control and treatment group. The treatment group would then receive encouragement while the control group remains ignored. However, such studies both require access to an appropriate platform and they also come with certain ethical concerns \cite{song:2010}. Pushed to the limit it would, for example, be unethical to withhold online social support from a suicidal person wanting to chat. Even not offering support to a person trying to lose weight could be questionable.

In this work, we propose a framework for conducting causation studies on weight loss from social media data. We make use of a growing collection of methods for causal inference from observational data. Though not without their limitations, such methods allow to go beyond arguing about correlations, attempting to rule out as many confounding factors as possible. In this study we look at the effect of receiving social support in a popular weight loss community, the /r/loseit subreddit, on users' self-reported weight loss. Specifically, we look at whether users who receive a certain number of comments on their first post in the community are more likely than those who do not to (i) return for another activity in the community, and to (ii) later report a higher weight loss, as measured to the community's badge system. To correct for content differences in users' posts, which are linked to receiving more or less support, we apply a matching approach: a post receiving a number of comments  bigger than a cutoff  is paired with a post very similar in content that received a number of comments smaller than the cutoff. Here similarity is defined in terms of posts exhibiting similar features derived via a statistical model using LDA topics \cite{Blei:2003}, Linguistic Inquiry and Word Count (LIWC)~\footnote{http://www.liwc.net} features, sentiment analysis \cite{Hutto:2014}, question-centric words and posts length.

Similar to previous work~\cite{cunha:2016, backstrom:2008}, we observe an increase in return probability to the community for those users receiving feedback. We then extend previous work by showing that, among returning users, those who had previously received comments on their post report higher weight loss than the matched control group who did not (46 lb vs.\ 37 lb). These findings are statistically significant.


To see if, for those users returning to the community, the difference in reported weight loss is mediated by a difference in (i) future lifetime in the community, or (ii) an increased engagement with the community we applied a so-called Sobel. Somewhat surprisingly, only about 5\% of the difference in reported weight loss appears to be mediated by an increased lifespan in the community, and this is not statistically significant even at $p=.1$. Instead, the \emph{rate of weight loss} is the main difference between the two groups.

Our main contributions are as follows.

\begin{itemize}
\itemsep-0.5em
\item We propose a framework to study the effect of receiving comments on a user's first post in a weight loss community on later reported weight loss.
\item Confirming prior work, we observed an increased return probability for those users who receive comments vs.\ those who do not.
\item Among those users who return and report weight loss through their badges, there is a 9 lbs difference between those who had previously received comments and those who had not.
\item We show that the difference in reported weight loss is not mediated by (i) a longer lifetime in the community or (ii) an increased activity level in the community.
\item We provide a detailed discussion of limitations, design implications and potential extensions.
\end{itemize}

We hope that the insights derived from our study lead to mechanisms further strengthening the social support online forums provide, especially to new users.

\section{Related works}\label{sec:related}

\textbf{On the importance of social support for positive health outcomes.}
Prior  research  has  extensively examined  the  role of social support in enhancing mental and physical health. It has been argued that receiving social support may reduce the rate at which individuals engage in risky behaviors, prevent negative appraisals, and increase treatment adherence \cite{fontana:1989}. Research has shown that conditions such as smoking \cite{mermelstein:1986}, depression \cite{grav:2012}, and coronary disease \cite{lett:2005} may be controlled with social support. Also, face-to-face support groups are positively correlated with desirable outcomes, such as lower blood pressure, and lower blood sugar levels, resulting indirectly from adaptive coping skills and responses \cite{sullivan:2003}. 

In the context of obesity, improvements in healthy eating and physical activity, as well more successful outcomes in weight reduction programs, have been demonstrated in studies considering offline support groups \cite{wang:2014}. 

\textbf{Online support forums.}
The main implication of these studies is that developing social support networks may help people manage their health conditions. Online health communities can be used to develop large social support networks, to understand and to promote health behavior. People have always tried to answer health related questions by themselves, now the Internet has become an important resource. Previous studies suggest that 30\% of U.S.\ Internet users have participated in medical or health-related  groups. Advantages of online communities include access to many peers with the same health concerns, and convenient communication spanning geographic distances. These communities present an interesting contrast to similar offline groups, as they provide an environment where people are more likely to discuss problems that they do not feel comfortable discussing face-to-face . In addition such online health communities are known to foster well-being, a sense of control, self-confidence and social interactions \cite{johnson:2006}. 

Still, little is known about how the support provided in these communities can help enhance positive health outcomes, such as weight loss. The literature offers little information about how members of large online health communities experience social support for weight loss. 
Most works concentrate on showing that social support exists in online weight-management communities, and qualify the types of support present online \cite{turner:2013,ballantine:2011}. For example, the presence of support was shown in popular weight-loss communities, including  SparkPeople~\cite{hwang:2010} and FatSecret~\cite{black:2010}. 

Based on the fact that support exists, a few studies have tried to correlate online engagement and support with the effectiveness of weight loss~\cite{turner:2013}, or to show that a network of engaged users is linked to persistent sharing of fitness related information~\cite{Park:2016}. The latter is of great importance as self-monitoring is one of the factors already shown to be associated with increased weight loss \cite{Hutchesson:2016}.
However, none of these studies were able to isolate the effects of online social support on weight loss, as there are many other underlying factors and real-world variables difficult to account for.
 
The main goal of our work is to move from detecting correlation and towards demonstrating causation. For that, we need to adequately control for covariates that affect the probability of receiving social support. For example, different types of people might differ in their ability to elicit social support because of differences in their personality, their mood, or their writing style. 
Similarly, mediation mechanisms that may underlie the observed relationship between social support and weight loss, and which are not typically investigated, need to be considered. 
We proposed to improve the ability to explain the effect of the support received by (i) applying a matching approach to reduce the bias in estimating the effect of receiving social support, and (ii) applying Sobel Tests to test for the existence of mediating effects. 



\textbf{Studies of health communities in Reddit.}\label{redditreview}
Reddit has been used to study different health conditions under different perspectives, including social support. For instance, Cunha {\it et al.}\ \cite{cunha:2016} study the r/loseit subreddit and observe that social support seems to be linked to an increase in return probability, a finding our analysis confirms.  
Nevertheless, the great majority of Reddit health community studies perform more exploratory analysis of users behavior or determine correlations between language and health outcomes, with the few studies looking at causality mentioned in the next subsection.
In the first category, 
Eschler \textit{et al.}\ \cite{eschler:2015} perform a content analysis in the posts of patients in different cancer stages in the subreddit r/cancer, showing that patients and survivor participants show different types of emotional needs according to their illness phase.
In the second category, Tamersoy \textit{et al.}\ \cite{tamersoy:2015} performed an analysis to identify key linguistic and interaction characteristics of short-term and long-term abstainers, focusing on tobacco or alcohol. 
 
\textbf{Causal inference from user-generated observational data.}
A common methodology for causal inference from observational data was borrowed from the domain of medicine and applies propensity score matching \cite{imai:2008}. 
This methodology is used by Choudhury \textit{et al.} \cite{de:2016} to identify suicidal ideation in Reddit mental health communities,  Tsapeli and Musolesi \cite{tsapeli:2015} to investigate the causal impact of several factors on stress level using smartphone data, Dos Reis and Cullota \cite{dos:2015} to study the effect of exercise on mental health, and Cheng \textit{et al.} \cite{cheng:2014} to understand whether community feedback regulates the quality and quantity of a user's future contributions in a way that benefits the community. Olteanu \textit{et al.} \cite{olteanu:2016} present a more general framework for applying the methodology to social media timelines.

However, we did not find any studies looking at causality between online variables and weight loss in online communities. 
Cunha {\it et al.}\ \cite{cunha:2016} apply a similar matching framework to the one proposed here to study the effect of social support on a change in return probability in the subreddit r/loseit. They do not, however, study the ultimately more important issue of \emph{weight loss}, and do not consider mediating effects. They also fail to report on formal measures of the quality of the matching, such as whether the covariates are indeed balanced in the control and the treatment group. 
Here, we apply this well-known methodology to study causality in weight loss, accounting for balance checking and mediation analysis. Our matching occurs directly on the variables and not on the propensity scores which, as shown in \cite{king:2015}, is preferable.





\section{Data}\label{sec:data}

Reddit\footnote{\url{https://www.reddit.com}} is a social news website and forum. Its content is organized in communities by areas of interest called subreddits. In 2015 it had 8.7 million users from 186 countries writing 73.2 million posts and 725.9 million comments in 88,700 active subreddits\footnote{``Active'' is determined by having 5 or more posts and comments during at least one week in 2015.}. For our study we look at the popular weight loss subreddit \emph{loseit}\footnote{loseit - Lose the Fat, \url{https://www.reddit.com/r/loseit/}.}.

The data used in our analysis covers five years (August 2010 to October 2014) and was crawled from Reddit using PRAW (Python Reddit API Wrapper), a Python package that allows simple access to Reddit's official API in November 2014. In Reddit users can submit content, such as textual posts or direct links to other sites, both collectively referred to as \emph{posts}. The community can then vote posted submissions up (\emph{upvotes}) or down (downvotes) to organize the posts and determine their position on the site's pages. Information on downvotes and upvotes is, however, not exposed via the API, instead they expose the aggregated number of votes, referred to as score (number of upvotes minus number of downvotes). Users can also reply to posts with \emph{comments}. 

The data we collected include posts, comments and other metadata (i.e., timestamp, user name, score). In total, we obtained 70,949 posts and 922,245 comments. These data were generated by 107,886 unique users, of which 38,981 (36.1\%) wrote at least one post and 101,003 (93.6\%) at least one comment. Table~\ref{tab:basicStatistics} shows the mean, median and standard deviation (SD) for basic statistics of the dataset, including the length of posts and comments and the number of daily messages. 

\begin{table}[ht]
\centering
\begin{tabular}{|l|l|l|l|}
\hline
							  & Mean & Median & SD \\ \hline
Posts per day &45.5 &45 & 22.7\\ \hline
Comments per day & 586.6 &599 & 264.3\\ \hline
Score per post & 35.7 &6 & 126.7\\ \hline
Score per comments &3.1 &2 & 11.4\\ \hline
Words per posts & 89.3 & 64 & 95.8\\ \hline
Words per comments & 25.5 & 14 & 35.3\\ \hline
\end{tabular}
\caption{Basic statistics of \textit{loseit} dataset.}
\label{tab:basicStatistics}
\end{table}

\begin{figure}[ht]
	\centerline{\includegraphics[width=0.45\textwidth]{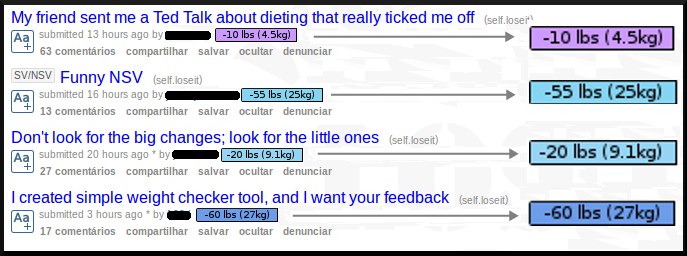}}
	\caption{Examples of the users' weight loss badges on loseit. The weight loss value is displayed in pounds and kilograms.}
	\label{fig:weightLossBadges}
\end{figure}

A participating user can add a ``badge'' (the icon which appears next to usernames, see Figure~\ref{fig:weightLossBadges}) to their profile that indicates self-reported information about their weight loss progress in pounds and kilograms. The badges can be updated by the users at any time. 

To answer our research questions we created two sets of users.

\textbf{Group 1 (G1).} We extracted the list of unique users whose first recorded activity in the community was a textual post (self post), rather than a comment or a post consisting exclusively of a URL (link post). This gave us a set of 25,647 users who had no public activity in the community prior to their post. We use this user set to study the effect of receiving comments on this post on the probability to return later. 

\begin{figure}[ht]	
\centerline{\includegraphics[width=0.45\textwidth]
	{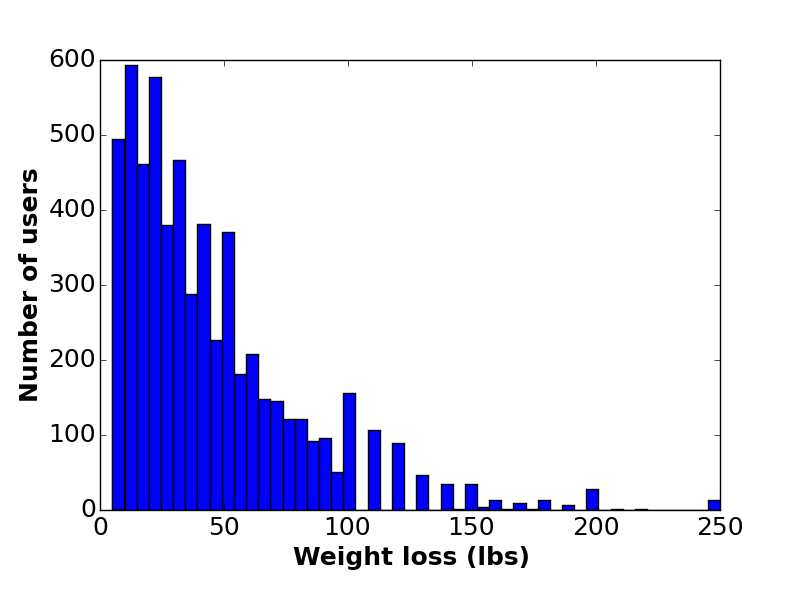}}
	\caption{Users' weight loss distribution displayed in the badges.}
	\label{fig:weightLossDistribution}
\end{figure}

\textbf{Group 2 (G2).} For users in Group 1, we extracted the list of unique users that both (i) returned again to the community later to comment or post and (ii) also had badge information indicating weight loss. This left us with a set of 6,143 users. Figure~\ref{fig:weightLossDistribution} shows the weight loss distribution displayed in the badges. We use this set of returning users to study the effect of receiving comments on their initial post on the weight loss they achieve. 

\section{methods}\label{sec:methods}

\begin{figure}[ht]	
\centerline{\includegraphics[width=0.45\textwidth]
	{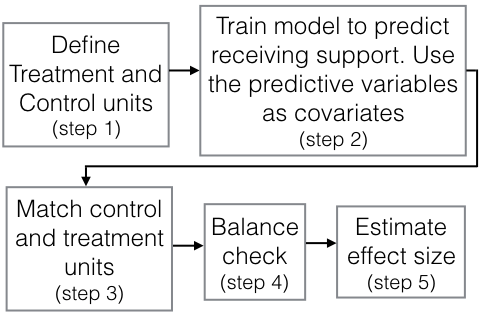}}
	\caption{Matching approach diagram.}
	\label{fig:framework}
\end{figure}

In this section, we discuss our matching methodology to investigate a potential causal effect of receiving social support in the Reddit loseit community. First we present the steps of our matching approach (see Figure~\ref{fig:framework}), later we explain the mediation test.


\textbf{Step 1 - Treatment and control definition -}  The first step of our matching approach is to choose the appropriate definition of treatment. Given the fact that 96\% of the first posts received at least one comment, we experimented with different definitions of treatment to avoid that the number of remaining matched observations becomes too small to draw any statistically significant conclusions. After running the experiments, we defined the treatment group as those users who received at least 4 comments on their first post in loseit. The control group consists of all users who received 3 or less comments on their first post. With this definition, we can guarantee (i) statistical significance of our findings, and (ii) the balance (see Step 4) between the two groups after performing matching.

\textbf{Step 2 - Statistical method for covariates selection -}  
Choosing appropriate confounding variables is an important step in matching methods. Ideally, conditional on the observed covariates, there should be no observed differences between the treatment and control groups. To satisfy the assumption of ignorable treatment assignment, it is important to include in the matching procedure all variables known to be related to the treatment assignment. Generally poor performance is found by methods using a relatively small set of ``predictors of convenience", such as gender only. Oppositely, including variables that are actually unassociated with the outcome can yield slight increases in variance. Commonly the confoundings' choice is based on previous research and scientific understanding, which can yield researcher discretion and bias \cite{stuart:2010}.

Here instead, we propose to use a statistical model to select the most important covariates. We first examine whether attributes of the content of posts, are predictive of receiving treatment. We model a prediction task with the data being split into two categories, the ones that received treatment and the ones that did not. Then we use the variables that remained in the final model as the confoundings in the matching approach (see Step 3).

In our case, the definition of the predictive variables was motivated by the hypothesis that posts with similar content have a similar probability of receiving feedback. Since user attributes like demographics or profile images are not available in Reddit, and hence the sole focus on the post's content is natural. We used a topical representation of the first post's content (title + body) extracted by Latent Dirichlet Allocation (LDA) \cite{Blei:2003}, the various semantic categories of words extracted from LIWC, sentiment analysis computed with Valence Aware Dictionary for sEntiment Reasoning (VADER) \cite{Hutto:2014}, counts of question-centric words (what, where, when, which, who, whose, why, how) and the length of a post (number of whitespace delimited words), a total of 78 variables. The required parameters for LDA -- number of topics, number of iterations, $\alpha$ and $\beta$ -- were empirically defined as 20, 2,000, 0.4 and 0.1. The rationale to get question words is to understand to what extent posts on weight loss seek explicit feedback or suggestions from the Reddit community.

We used a logistic regression with LASSO as our prediction method. Logistic regression is well-suited to handle binary dependent variables, while LASSO is a method that performs both variable selection and regularization in order to enhance the prediction accuracy and interpretability of the statistical model. To assess the quality of the model produced, we (i) computed the mean AUC (Area Under the Curve) over the 10-fold cross validation setting, and (ii) performed a qualitative analysis using the features that remained in the model to compute the similarity between posts. As the goal of the matching is to pair posts that ``look similar to a human reader'', this qualitative analysis is important to understand whether features have succeeded to adequately identify similar texts.

The highest cross-validated AUC of 0.62 was obtained for 37 variables. However, in order to further reduce the dimensionality of the space used for matching, we chose features from a model with slightly lower AUC (0.61) but which used only 20 variables. In terms of qualitative analysis, both models fared similarly without any noticeable difference.

The 20 variables that ``survived'' the shrinkage were: 5 LIWC categories (negative emotions, anger, sexual, reward and work), sentiment and 14 LDA topics. Additionally we use the coefficient values as covariates ``weights'' in the similarity computation in Step 3.  This choice was motivated by the fact that the regression coefficients have two desirable properties. The first one is a scale normalization property, where something measured in, say, kilometers would have a larger coefficient than the same property measured in meters. This normalization is crucial for computing meaningful similarities in a metric space. Second, they reflect an importance of the predictive variable in relation to the response variable. This means that variables with more effect on receiving feedback will be given more importance on the post similarity.

\textbf{Step 3 - Matching approach - }Matching is a nonparametric method of controlling for the confounding influence of pretreatment control variables (also known as confounding or covariates) in observational data. The key goal of matching is to \emph{prune observations} from the data so that the remaining data have better balance between the treated and control groups, meaning that the empirical distributions of the covariates in the groups are more similar and model dependency is reduced \cite{king:2014}.

Without matching we may have imbalance, for example, a generally optimistic user might write a first post with a more positive tone than a more pessimistic counterpart. Let us imagine that, in response to their posts, the former user receives lots of support and the latter receives none. Now let us further imagine that the former user returns for more activity on the subreddit later, whereas the latter user is never to be seen again. The question then arises whether the support received ``caused'' the former user to return or, rather, whether that user was at a higher disposition to return anyway and the social support received was a mere correlate. Here the tone of the posts, an important covariate, is imbalanced and is generally more positive in the treated group (= those with social support) than in the control group (= those without social support). Matching approaches are applied in such scenarios to remove the relationship between the covariates and the supposed causal variable by reducing the imbalance.



In the simplest case, matching is applied to settings of a dependent outcome variable $Y_{i}$, a treatment variable $T_{i} (1 = treated, 0 = control)$ and a set of pretreatment covariates $X_{i}$ \cite{rubin:1976}. We want to observe the treatment effect for the treated observation \textit{i} ($TE_{i}$), which is define as the value of \textit{i} when \textit{i} receives the treatment minus the value of \textit{i} when it does not receive the treatment.

\begin{equation}
		TE_{i} = Y_{i}(1) - Y_{i}(0) \\
		    = observed - unobserved
\end{equation}

Obviously if \textit{i} is treated, we can not also observe \textit{i} when it does not receive the treatment. Hence, matching estimates $Y_{i}$(0) with a $Y_{j}$(0), where \textit{j} is similar to \textit{i}. In the best case, each \textit{i} is \emph{matched} to a \textit{j} with the exact same values for all the control variables. In practice, ``similar enough'' observations are being matched. 


Matching can be viewed as trying to find hidden randomized experiments inside observational data. The most commonly used matching method is Propensity Score Matching (PSM) \cite{king:2015}, which aims to approximate a complete random experiment. PSM first builds a model to predict the probability of a particular user to receive the treatment. Users are then matched according to their probability of receiving the treatment. However, recently King and Nielsen \cite{king:2015} showed that this method is suboptimal and that PSM can, under certain circumstances, even increase the bias in the data.

Here, we apply a matching distance approach (MDA) \cite{rubin:2006}, which aims to approximate a fully blocked experiment \cite{imai:2008}. For this we measure the cosine distance among the observations based on their covariates. Treated units are matched to their nearest control, assuming they pass a predefined similarity threshold, a.k.a.\ caliper. Ideally this similarity threshold should be as close as possible to 1, barring constraints related to data sparsity. To find an appropriate value, we gradually increase the value, starting from 0.9, until we are able to observe three conditions: (i) the matched posts are similar enough (based on a qualitative analysis), (ii) treatment and control groups are balanced (see Step 4), and (iii) results are statistically significant. We allow one-to-many matches, i.e., we match with replacement.

Pruning the unmatched observations makes the control variables matter less. In other words, it breaks the link between the confounding and the treatment variable, consequently reducing the imbalance, model dependence, research discretion and bias.

\textbf{Step 4 - Balance check -} One necessary condition for a successful application of a matching methodology is a balance of the co-variates. If, say, one LDA topic was more strongly pronounced in the treatment group than in the control group then this imbalance, rather than any causal effect, could lead to an apparent treatment effect. To assess if the treatment and control groups are sufficiently balanced after the matching, we check the standardized mean difference \cite{austin:2011} for each confounding variable \textit{c}. For a continuous covariate, the standardized mean difference is defined as:

\begin{equation}
		d_{c} = {(\bar{x}_{treatment} - \bar{x}_{control}) \over \sqrt{s^{2}_{treatment}+ s^{2}_{control} \over 2 } }
\end{equation}

where $\bar{x}_{treatment}$ and $\bar{x}_{control}$ denote the mean of the covariate in the treatment and control groups, respectively. $s^2_{treatment}$ and $s^2_{control}$ denote the corresponding sample variances.

The standardized difference compares the difference in means in units of the pooled standard deviation. It is not influenced by sample size and allows for the comparison of the relative balance. The remaining bias from a confounding variable \textit{c} is considered to be insignificant if $d_c$ is smaller than $0.1$ \cite{mimno:2011}.

\textbf{Step 5 - Effect size estimation -} After showing that any confounding bias has been sufficiently eliminated, we can estimate the effect of  treatment on the matched treated and control units. Here for a given matching of treated and control units, we compute the estimated average treatment effect (EATE).

\begin{equation}
		EATE = {\sum_{i=1,j=1}^{N} {(Y_{i}(1) - Y_{j}(0))*100 \over Y_{j}(0)} \over N}
\end{equation}

\textbf{Mediation test - }Though matching methods can shed light on whether a change in the treatment condition T likely causes a change in the dependent variable Y, matching methods do not provide insights into whether (i) this causal relationship is ``direct'', or whether (ii) it is being mediated by another variable M. In our case, receiving social support might lead to an increased engagement with the community which, in turn, is responsible for an increase in weight loss success. Thus the weight loss success would be mediated by an increased engagement with the community.

Mediation analysis is the process of determining whether or not variables acting as an in-between step, called mediators, are present when looking at the relationship between an independent variable T (here the treatment confidtion) and a dependent outcome variable Y. As a result, when the mediator is included in an analysis model with the independent variable, the effect of the independent variable is reduced and the effect of the mediator remains significant \cite{boone:2012}.

To verify if any variable plays the role of a mediator and its significance in the relationship of social support and weight loss, we apply the Sobel test \cite{sobel:1986}. The Sobel test assesses the statistical significance of the indirect effect.


Note that such analysis works naturally with matching as a preprocessing step: the matching reduces imbalance between the treated and control groups in terms of the covariates used for matching. Hence, the remaining unpruned observations are similar except for the treatment condition, and the treatment condition can be used as an independent variable in the Sobel test.

\section{Results}\label{sec:results}

In this section we present the results of the causal inference analysis we conducted to measure the effect of receiving social support for the first post a user shares with the community. In Figure~\ref{fig:effectSize} the stars indicate the significance levels for a permutation test, with the number of asterisks corresponding to the \textit{p}-values, *** for 0.1\%, ** for 1\%, and * for 5\%. In a permutation test, the labels (C)ontrol and (T)reatment are repeatedly randomly shuffled and for each (fake) control-treatment assignment the effect size is measured. The significance level indicates the fraction of permutations that lead to an effect size bigger than the one actually observed.\cite{hesterberg:2005}.


\subsection*{User engagement}

We start by analyzing the effect of receiving social support on the probability of a user to return to the community to comment or post again. For this analysis we used the 25,647 users present in \textbf{Group 1} (see Section~\ref{sec:data}). Among those users, 18,000 received the treatment (at least 4 comments) and 7,647 did not (control).

\begin{figure}[ht]	
\centerline{\includegraphics[width=0.45\textwidth]
	{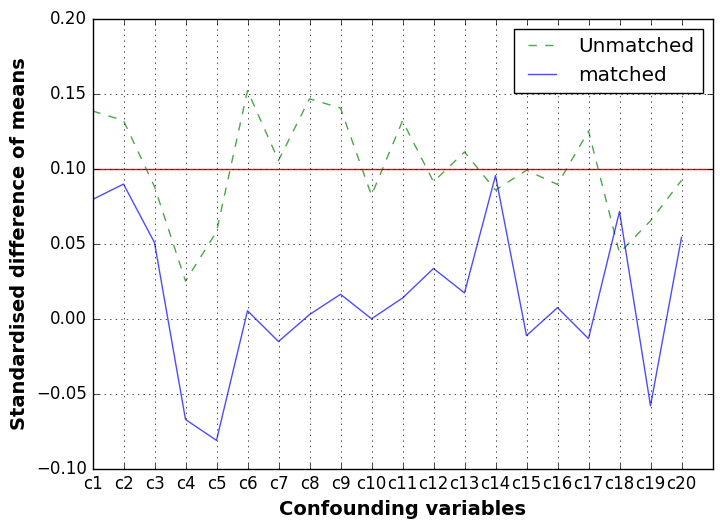}}
		\caption{Standardized difference of means for each confounding variable (Group 1). Note that after the matching all the values are below 0.1, thus the groups are balanced.}
	\label{fig:balanceReturn}
\end{figure}

\begin{figure}[ht]	
\centerline{\includegraphics[width=0.45\textwidth]
	{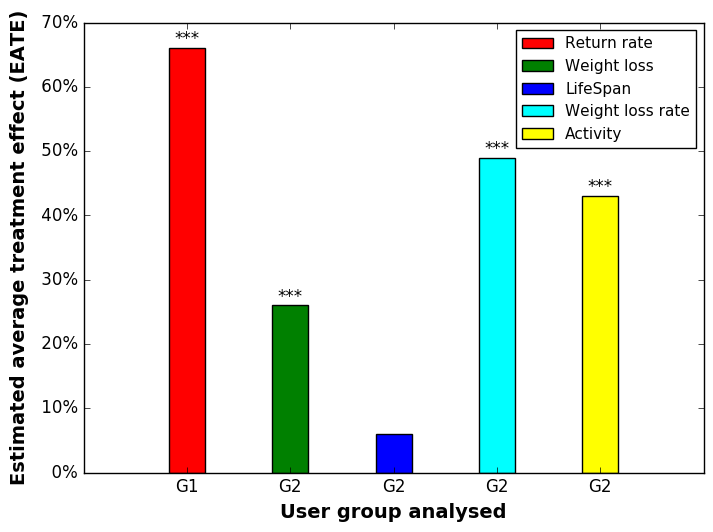}}
		\caption{The effect size for the factors analyzed.}
	\label{fig:effectSize}
\end{figure}

We applied our one-to-many matching approach with a similarity threshold of 0.965 to ensure that for \textbf{Group 1} our method was matching similar enough posts and balancing the groups (see Figure~\ref{fig:balanceReturn}). The matching produced 14,570 similar pairs (14,570 unique treatment and 5,279 unique control users). Our results indicate that receiving social support increases the relative probability of a user returning to the community by roughly 66\% (see Figure~\ref{fig:effectSize}, red bar). This analysis is statistically significant at 0.1\%. These results provide evidence for how important social support is to increase user engagement, which is associated with better chances of obtain success in weight loss programs \cite{patrick:2011}. Note that this effect sizes is bigger than the one reported in a similar study \cite{cunha:2016}. There the authors used a different definition of control and treatment group based on ranking the posts by the number of comments received than getting ``top 40\%'' vs.\ ``bottom 40\%'' comments, rather than our ``at least 4'' vs.\ ``at most 3''. 


%
%

To show that the matching approach indeed matches similar posts, we present in Table~\ref{tab:similarPosts} parts of a pair of posts matched according to our approach, this pair had a cosine similarity of 0.97.

\begin{table}[ht]
\centering
\resizebox{\columnwidth}{!}{%
\begin{tabular}{|p{7.9cm}|}
\hline
\textbf{Treatment:} Hi guys, new here, I'm on a low carb and dairy (pale) diet, but recently i just went on a vacation, I ate..- Wrap with grilled chicken, lettuce, and a small amount of buffalo sauce, on whole grain wrap,Banana, Apple, Orange, and plain oatmeal - Tossed salad with grilled chicken. Is this healthy eating on the pale diet? I also did not exercise, but we did some walking around...\\ \hline
\textbf{Control:} Hey everyone, I'm new to loseit.  I'm starting my first workout/diet routine ever. My doctor says I need to get my cholesterol under control.  So far, I've been doing 30 minutes of cardio  and taking care of my diet. I've cut out soda, beer, and red meat. I've switched to skim milk, olive oil, whole grain, and brown rice...\\ \hline
\end{tabular}}
\caption{Parts of similar posts matched.}
\label{tab:similarPosts}
\end{table}

\subsection*{Weight loss}

\begin{figure}[ht]	
\centerline{\includegraphics[width=0.45\textwidth]
	{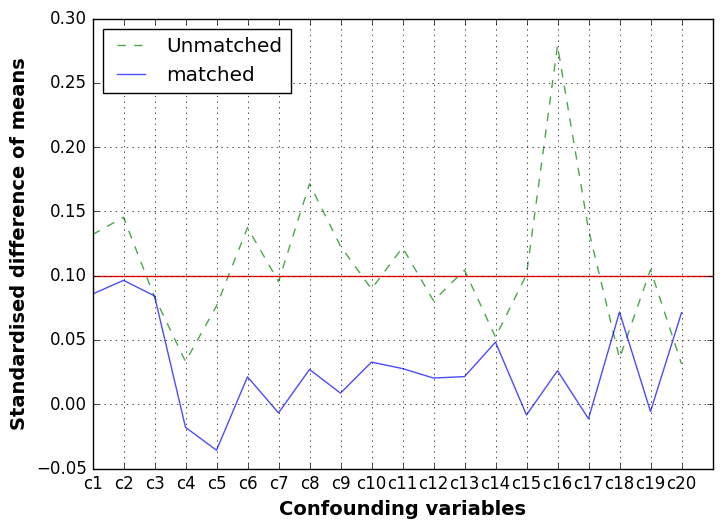}}
	\caption{Standardized difference of means for each confounding variable (Group 2). Note that after the matching all the values are below 0.1, thus the groups are balanced.}
	\label{fig:balanceWeightLoss}
\end{figure}

Next we investigated the effect of receiving social support on weight loss. For this analysis we used the set of 6,143 users in \textbf{Group 2} (see Section~\ref{sec:data}), among those users 4,657 received at least 4 comments (treatment group) and 1,486 did not (control group). Here, to ensure similar enough posts and balanced groups in the matching for \textbf{Group 2}, we used a  similarity threshold of 0.955. Figure~\ref{fig:effectSize} (green bar) shows that receiving social support in the first post leads to a relative increase in the achieved weight loss of 26\%, or an absolute mean difference of 9 lbs. The observed effect size is statistically significant and after applying the balance check (see Figure~\ref{fig:balanceWeightLoss}) we confirmed that all covariates were balanced between the control and treatment groups.

However, note that the frequency with which people update their badges may interfere in this analysis. Maybe users who do not get comments do not update their badges as often, even if they lose as much weight as others. In other worse, receiving social feedback might simply lead to more active ``profile management'' than to more weight loss. To test this alternative explanation, we computed the number of badge updates (every change in the badge information) for users in the treatment and control groups. Afterwards, we ran a permutation test to check if the two groups' badge updating behaviors were similar. The groups presented a mean number of updates of $1.75 \pm 3.93$ (treatment) and $1.60 \pm 2.70 $ (control), but this difference was not statistically significant, (\textit{p}=0.16). As the two groups' badge updating behaviors are similar and it does not seem to effect our analysis.

\begin{figure}[ht]	
\centerline{\includegraphics[width=0.45\textwidth]
	{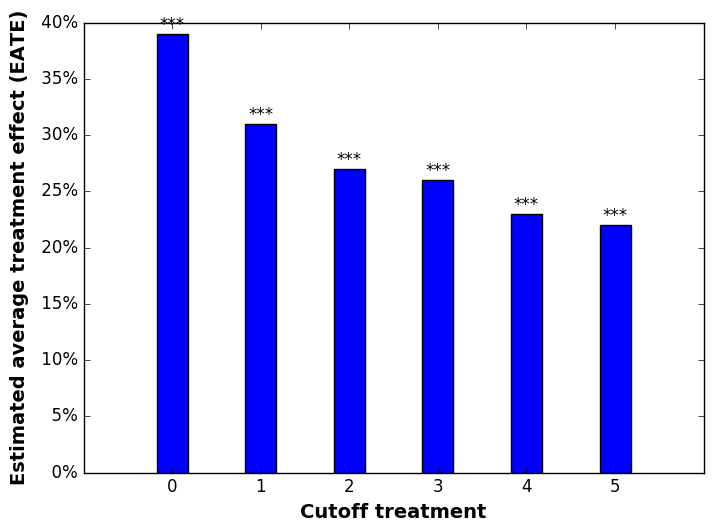}}
	\caption{Effect size for different treatment definitions.}
	\label{fig:cutoffs}
\end{figure}

We also experimented with different definitions for the treatment cutoff to see if there is an effect of ``diminishing returns'': receiving at least one comment (vs.\ none) could have a bigger impact than receiving 10 (vs.\ 9 or less). Figure~\ref{fig:cutoffs} presents the effect size for different treatment definitions, although we can not guarantee the balance property for all the cutoffs, these results show that as expected when we increase the cutoff the effect size drops.

\subsection*{Mediation test}

After estimating the causal effect of receiving social support on weight loss, we focus on checking if certain variables that were not included in the set of covariates could act as a mediator, explaining part of the observed effect of social support on weight loss. Conceptually and based on prior work, receiving social feedback could cause the effected user to (i) show a higher activity level in the community, and (ii) remain longer in the community. Therefore we first check if, indeed, receiving a comment on a user's first post has an effect on these variables and, if yes, if this effect mediates the observed effect on the reported weight loss.

Figure~\ref{fig:effectSize} shows that receiving social support also has an effect over the users' lifespan (i.e., the difference in days between the date of their last and first activity observed in the community) and the of number activities (i.e., the sum of the number of comments and posts). The effect size for lifeSpan is 6\% (see Figure~\ref{fig:effectSize}; blue bar), but this effect was not statistically significant, for the number of activities the effect size was 43\% (see Figure~\ref{fig:effectSize}; yellow bar).

Since the observed effect on the lifespan is small, we estimated if the social support has an effect on the users' weight loss rate (i.e., the weight loss in lb divided by the lifespan). As the rate is an unstable estimate for users who are only active for one or two days in the community, we chose to look at the median rather than the average effect. Concretely, we computed the median of the individual paired ratios of (weight loss rate treated individual / weight loss rate control individual). We then use the median of these medians as an estimate of the effect size. As expected there is an effect on the weight loss rate, where users that received at least 4 comments (treatment) lose weight roughly 35\% faster than the ones that did not received (0.48 lb/day vs.\ 0.35 lb/day).

Finally, we applied a Sobel test to verify if lifespan and number of activities act as mediator in the relationship of the social support and weight loss, i.e., if they explain a significant part of the causal effect of social support on weight loss. The results of the Sobel test showed that the proportion of the effect of social support over weight loss due to lifespan and the number of activity is small -- 5.6\% and 3.4\% respectively. However, these results were not statistically significant even at $p = 0.1$. Surprisingly, among the users that return to post again, the difference in achieved weight loss does not seem to be linked to either lifespan or engagement in the community. Rather, the \emph{rate of weight loss} seems to be effected for those users returning to the community.

\section{Discussion}\label{sec:discussion}

\textbf{Does weight loss equal success?} Our analysis crucially assumes that members of the loseit community \emph{want} to lose weight. If that was not the case then talking about ``weight loss success'' would be meaningless. However, results from a recent survey of the loseit community~\cite{survey} indicate that 91\% of the respondents were currently trying to lose weight, with another 7\% trying to maintain their weight. Therefore, it seems adequate to consider a higher level of weight loss as a desirable outcome.

\textbf{Qualitative evidence.} Though our analysis is deliberately using quantitative methods, there is also qualitative evidence to further support the claim that social support received in the community effects weight loss. 
The aforementioned community survey~\cite{survey} includes the question ``What do you like about /r/loseit?''. The topics most emphasized by the survey participants were related to terms such as ``community'', ``people'', ``supportive'' and ``support''. Similarly, one can easily find posts explicitly acknowledging the perceived importance of the social support such as: ``I have visited this page almost daily over the past 15 months, and it was really helpful in keeping my motivation. I hope this may provide similar motivation to those just starting! '' .


\textbf{Designs implications.} Our findings suggest that whether or not a user receives feedback on their initial post affects (i) their probability to return to the community, and (ii) given that they return, the amount of weight loss they report. Therefore mechanisms that increase the likelihood to receive a ``warm welcome'' are expected to lead to more engagement with the community and to better health outcomes. Fortunately, the vast majority of initial posts (96\%) already receive at least one comment. It could be worth considering a mechanism where posts that do not are brought to the moderators' attention so that they can provide an adequate reply. It could also be worthwhile to construct a ``positivity bot'' which provides non-generic positive feedback on posts overlooked by the community\cite{vanderzwaan:2012}. Other researchers are exploring the creation of a framework to allow formal testing of theories of different moderation styles.\footnote{See~\url{https://civic.mit.edu/blog/natematias/reddit-moderators-lets-test-theories-of-moderation-together} for thoughts by Nathan Matias' and \url{https://np.reddit.com/r/TheoryOfReddit/comments/456503/want_to_test_your_theories_of_moderation_lets/} for a subreddit on the topic.}  Our research contributes by providing a theory to test.


\textbf{Ethical considerations.}
For our study, we used only publicly available data that users chose to post online. All analysis is done in aggregate and we do not post results for any individual.  However, as is often the case with such data collection, users might not be aware of the fact that they are being studied by researchers. To at least partly alleviate such concerns, we reached out the moderators to inform them about our study. Their reaction was very positive (``Wow, I'm really looking forward to it'') and they also pointed us to the community survey \cite{survey} that we had previously been unaware of. Once finalized, we will share our findings with the loseit community to encourage a positive atmosphere and, in particular, ensure a warm welcome of new members.

\textbf{What type of social support matters?} One could also try and extract the topics or tone from the comments on a given post to see if particular types of comments have a larger effect on the reported weight loss. This, however, comes with endogeneity problems as the type of comments received is likely correlate with the subject matter of the post. Given large enough data sets one could hope to correct for this using our matching framework where the treatment is no longer binary -- receiving a comment or not -- but is multi-variate. We chose not to explore this route due to sparsity concerns.

We did, however, experiment with using another type of social feedback based on votes: Reddit has a voting system with up- and down-votes and an aggregate ``score'' combining these two types of votes, positive - negative, can be obtained via the API. In our data set, this score was never negative. Using the same matching setup (see Section~\ref{sec:methods}), this gave an effect size of 16\% to users to return to the community and 45\% to weight loss ($p<0.01$) . The balance condition also held for this experiment. Though most of the limitations discussed below also apply to this setup, the fact that a ``similar'' effect is observed for a different definition of social feedback indicates that our results might hold more general.

\textbf{Who benefits most from social support?} All of our results are aggregates indicating that, on average, users seem to benefit from receiving social support in the form of comments. For future work it would be interesting to look into what type of users are most or least likely to benefit from such support, for example looking for gender-specific effects. Though gender is not an attribute of a user's profile, it can sometimes be inferred from their posts (``I'm a mother of two ...'') or from shared progress pictures. In some cases a user's chosen screen name such as ``john123'' or ``mary456'' also provide hints. Similarly, one could look for cases of users indicating their starting weight, rather than just the weight loss, to study whether the effect of social support is tied to a user's initial weight.

\textbf{Impact of pruning on effect size.} We started our analysis with the assumption that, as we had previously observed for the effect on return-to-post probability \cite{cunha:2016}, pruning via matching would lead to a \emph{lower} estimate of the effect of social feedback on weight loss when compared to the unmatched analysis. Intuitively, matching, and hence pruning, should reduce the effect of confounding user variables such as ``positive outlook on life'' which might affect writing style and have a positive effect on both return-to-post behavior as well as on weight loss success. However, we observed the \emph{opposite}: the raw effect size for the unmatched data was 22\% (treatment cutoff of 4 comments) whereas it was 26\% for the matched analysis (treatment cutoff of 4 comments and similarity threshold of 0.955). In particular, users who were treated and who, eventually, lost less than the median weight loss were pruned more often (27\%) than their treated counter-parts who lost a lot of weight (23\%). At the same time for the untreated users the differences in pruning rates for less-than-median (28\%) and more-than-median (28\%) weight loss were small.  Though this is surprising, it actually helps to make the overall claim, i.e., that social feedback supports weight loss in an online community, more robust.

\section{Limitations}

\textbf{Limitations of using badges to track weight loss.}
To infer a user's weight loss success or failure we are currently relying on the badges used in the loseit community. These badges only capture \emph{self-reported} weight loss progress. The first issue imposes an important limitation as, one could imagine, receiving social feedback leads to a heightened sense of self-awareness and a feeling of ``being watched'' in the community. Though this could lead to positive peer pressure, it could also increase the probability of over-reporting weight loss progress. Badges are also always visibly displayed next to a user's screen name, increasing the likelihood of social signaling effecting their use. One solution to this issue could be to perform a similar analysis using auto-generated weight updates from smart scales as used by Wang {\it et al.} \cite{wang:2016}. Such data sources are less likely to be prone to misreporting errors.

To avoid at least some of the limitations introduced by using the badges to infer a user's weight loss, we also performed analysis on a different, smaller set of users who explicitly report their weight loss progress in their posts. This can be either by using the community conventions of ``SW/CW/GW'' for starting/current/goal weight, or through posting things such as ``I've lost 10lbs''. In this analysis the treatment group had 1,421 users and the control group 36 users. Due to data sparsity, we only applied the threshold of 0.7. Though prone to other limitations, this alternative way of inferring weight loss led to qualitatively similar effect sizes of 9.5\%.

\textbf{Limitations to determining the start date of weight loss journey.}
Our analysis, especially that related to the rate of weight loss (Section~\ref{sec:results}), assumes that a user's weight loss journey starts the day they first announce themselves to the community in the form of a post. However, in practice, users might well first observe the community before deciding to post. This means that the actual rate of weight loss is likely to be lower, as the time period over which the weight loss is achieved is longer. A more subtle issue related to this passive use of the support community is that it could affect the writing style. Put simply, users who have been following the community for a while might (i) have a ``head start'' as far as weight loss is concerned, and (ii) they might write in a style more in tune with the community which, in turn, could lead to more social feedback. If these stylistic differences are not represented in the extracted covariates, this could lead to an overestimate of the effect size. Though we cannot completely rule out this possibility, we believe that the effect sizes are large enough to make it unlikely that they are fully explained by this hidden adaption to the community.


\textbf{Limitations of our matching approach.} 
When applying a matching approach, there are a number of choice one needs to make such as (i) the selection of covariates, (ii) how to normalize and potentially weight different covariates, (iii) which distance metric to use and whether to use ``blocking'', and (iv) which similarity threshold to choose. Of these, any choices for (ii), (iii) and (iv) should asymptotically converge to the same result as the matched pairs become more and more similar, being identical on all the covariates in the limiting case. We therefore did not include experiments with higher similarity thresholds because of issues of data sparsity.

Concerning the covariates used, we believe our statistical method is a reasonable choice. Introducing too many additional covariates can lead to problems of high dimensionality when attempting to match similar posts along, say, hundreds of dimensions. In such settings it becomes difficult to balance all the covariates considered. Furthermore, many other potential covariates should be balanced at least ``on average'', if they are correlated with the covariates in the final model. Still, the exclusion of unknown but potentially crucial covariates is always a concern when applying a matching methodology. We hope that other researchers will validate -- or invalidate -- our analysis using their own set of covariates. 

\textbf{Limitations of observability of returning users.}
Our main analysis relies on users returning to the loseit community to post or comment again after their initial post. Assuming the user also has badge information, then this second data point provides an estimate of both the absolute weight loss (in the badge) and the weight loss rate (in the difference of time stamps). This means that we cannot make any statements concerning weight loss for users who do return to the community for a second, public activity. Though our main results are conditional on users returning, we also observe social feedback leading to an increase in return probability (see Figure~\ref{fig:effectSize}). However, the issue of predicting who will or will not return to an online community has been studied before \cite{cunha:2016,Choudhury:2014} and so we did not analyze this further.

\section{Conclusions}\label{sec:conclusions}

We presented an analysis of the effect of receiving social feedback in the form of comments on the reported weight loss in the /r/loseit community on Reddit. Correcting for confounding factors through a matching methodology, users who receive at least 4 comments on their first post in the community were (i) 66\% more likely to return for a future activity in the community, and (ii) conditional on the user returning, those who had previously received feedback end up reporting on average 9 lb more in weight loss. For these returning users, this effect is not mediated by neither (i) an increased level of activity in the community, nor by (ii) a longer lifespan in the community. Though observational studies have inherent limitations on causal inference, our work helps to illustrate the importance of receiving feedback in online support forums, in particular for users new to the community.

%
\bibliographystyle{abbrv}
\bibliography{refs}  
%
%

\end{document}